\documentclass[a4paper,twocolumn,11pt,accepted=2023-04-10]{quantumarticle}
\pdfoutput=1
\usepackage[utf8]{inputenc}
\usepackage[english]{babel}
\usepackage[T1]{fontenc}
\usepackage{amsmath}
\usepackage{hyperref}

\usepackage{tikz}
\usepackage{lipsum}

\usepackage[numbers]{natbib}

\usepackage{amssymb,amstext,amsmath,bm,bbm}
\usepackage{xcolor}

\usepackage{tikz-cd}
\usepackage{hyperref}
\usetikzlibrary{positioning}
\usetikzlibrary{decorations.markings}

\usepackage[shortcuts]{extdash}

\newcommand{\Dj}{\mbox{\raise0.3ex\hbox{-}\kern-0.4em D}}
\renewcommand{\dj}{d\kern-0.4em\char"16\kern-0.1em}

\newcommand{\bra}[1]{{\langle{#1}\vert}}
\newcommand{\ket}[1]{{\vert{#1}\rangle}}
\newcommand{\bbra}[1]{{\langle\!\langle{#1}\vert}}
\newcommand{\kket}[1]{{\vert{#1}\rangle\!\rangle}}
\newcommand{\bracket}[2]{\langle #1 \vert #2 \rangle}

\newcommand{\bbrakket}[2]{{\langle\!\langle{#1}\vert {#2} \rangle\!\rangle}}

\newcommand{\tr}{\mathop{\rm Tr}\nolimits}

\newcommand{\prirodni}{\ensuremath{\mathbb{N}}}

\newcommand{\kompleksni}{\ensuremath{\mathbb{C}}}

\newcommand{\sspan}{\mathop{\rm span}\nolimits}

\newcommand{\one}{\mathbbm{1}}

\newcommand{\ds}{\displaystyle}

\newcommand{\cF}{{\cal F}}
\newcommand{\cG}{{\cal G}}
\newcommand{\cH}{{\cal H}}
\newcommand{\cI}{{\cal I}}

\newcommand{\cM}{{\cal M}}

\newcommand{\cP}{{\cal P}}
\newcommand{\cW}{{\cal W}}

\begin{document}

\title{Operational interpretation of the vacuum and process matrices for identical particles}

\author{Ricardo Faleiro}
\affiliation{Instituto de Telecomunica\c{c}\~oes, Avenida Rovisco Pais 1, 1049-001, Lisboa, Portugal}
\email{ricardofaleiro@tecnico.ulisboa.pt}
\orcid{0000-0002-4155-7396}

\author{Nikola Paunkovi\'c}
\affiliation{Instituto de Telecomunica\c{c}\~oes, Avenida Rovisco Pais 1, 1049-001, Lisboa, Portugal}
\affiliation{Departamento de Matem\'atica, Instituto Superior T\'ecnico, Universidade de Lisboa, Avenida Rovisco Pais 1, 1049-001, Lisboa, Portugal}
\email{npaunkov@math.tecnico.ulisboa.pt}
\orcid{0000-0002-9345-4321}

\author{Marko Vojinovi\'c}
\affiliation{Institute of Physics, University of Belgrade, Pregrevica 118, 11080 Belgrade, Serbia}
\email{vmarko@ipb.ac.rs}
\orcid{0000-0001-6977-4870}


\maketitle

\begin{abstract}
This work overviews the single-particle two-way communication protocol recently introduced by del Santo and Daki\'c (dSD), and analyses it using the process matrix formalism. We give a detailed account of the importance and the operational meaning of the interaction of an agent with the vacuum --- in particular its role in the process matrix description. Our analysis shows that the interaction with the vacuum should be treated as an operation, on equal footing with all other interactions. This raises the issue of counting such operations in an operational manner. Motivated by this analysis, we apply the process matrix formalism to capped Fock spaces using the framework of second quantisation, in order to characterise protocols with an indefinite number of identical particles.
\end{abstract}

\section{\label{sec:intro}Introduction}

In recent years there have been advances in quantum information theory related to new techniques for discussing quantum circuits and quantum computation. One of those techniques is the recently developed process matrix formalism~\cite{ore:cos:bru:12}. This formalism is general enough to describe all known quantum processes, in particular the superposed orders of operations in a  quantum circuit. Moreover, its most prominent feature is that it allows for a description of more general situations of indefinite causal orders of spacetime points. A formal example of such a process has been introduced and discussed in~\cite{ore:cos:bru:12}, leading to the violation of the so-called {\em causal inequalities}. The latter represent device-independent conditions that need to be satisfied in order for a given process to have a well-defined causal order. It is still an open question whether such a process is physical and can be realised in nature. Also, a lot of attention in the literature has been devoted to the {\em quantum switch} operation, which has been discussed through both theoretical descriptions~\cite{chi:dar:per:val:13,ara:bra:cos:fei:gia:bru:15,bav:ara:bru:que:19,gue:rub:bru:18} and experimental implementations~\cite{pro:etal:15,rub:roz:fei:ara:zeu:pro:bru:wal:17,rub:roz:mas:ara:zyc:bru:wal:22}.

One of the interesting aspects of the quantum switch is that it gives rise to a superposition of orders of quantum operations. In a recent work~\cite{pau:voj:20}, the difference between the superposition of orders of quantum operations and the superposition of causal orders in spacetime was discussed in detail, and it was demonstrated that the latter can in principle be realised only in the context of quantum gravity (see also~\cite{vil:ren:22,orm:van:bar:22,vil:17}). The detailed analysis of the causal structure of the quantum switch has revealed one important qualitative aspect of the process matrix description --- in order to properly account for the causal structure of an arbitrary process, it is {\em necessary} to introduce the notion of the quantum vacuum as a possible physical state. Otherwise, the naive application of the process matrix formalism may suggest a misleading conclusion that quantum switch implementations in flat spacetime feature genuine superpositions of spacetime causal orders. This demonstrates the importance of the concept of vacuum in quantum information processing. Regarding the general role of the vacuum in quantum circuits and optical experiments, see \cite{abe:04} and \cite{chi:kri:19, kri:mao:chi:21}, respectively, and the references therein.

Simultaneously with these developments, another interesting quantum process has been recently proposed~\cite{dak:san:18} by del Santo and Daki\'c  --- dSD protocol (see also subsequent theoretical \cite{san:dak:20, gho:muk:ara:19} and experimental work \cite{mas:moq:bau:san:ket:dak:wal:19}). As it turns out, while this process enables Alice and Bob to guess each other's input bits with certainty by exchanging a single particle only once, it cannot be correctly described within the process matrix formalism without the introduction of the interaction between the vacuum and the apparatus as an {\em operation}. Thus, it represents an additional motivation to introduce the vacuum state into the process matrix formalism, independent of the reasons related to the description of the quantum switch process.

Moreover, while the dSD protocol employs only one photon, it is also relevant for multiphoton processes, which opens the question of the treatment of identical particles within the process matrix formalism. Also, taking into account the presence of the vacuum state, one is steered towards the application of the abstract process matrix formalism to systems with variable number of identical particles, to the second quantisation and ultimately generalisation to quantum field theory (QFT). See also a related work on the causal boxes framework~\cite{por:mat:mau:ren:tac}.

In this work we give a detailed description and treatment of dSD protocol within the process matrix formalism. We analyse in detail the role of the vacuum in the protocol and the formalism, and its operational interpretation. Specifically, our aim is to discuss the following question:

\begin{center}
{\em Is the interaction with the vacuum an operation, or not?}
\end{center}

\noindent Our analysis of dSD protocol leads to a conclusion that the interaction with the vacuum should be considered an operation. The alternative would mean that one could extract information from the system at the final moment of the protocol without performing an operation at all. Since the same physical situation should always be described in the same way, we conclude that the interaction with the vacuum should be treated as an operation, and thus as a resource, in all quantum information protocols. This includes the optical implementation of the quantum switch protocol, leading one to infer that it features {\em four, rather than just two} operations, as was claimed in a number of papers \cite{ara:bra:cos:fei:gia:bru:15, bav:ara:bru:que:19, gue:rub:bru:18, pro:etal:15, rub:roz:fei:ara:zeu:pro:bru:wal:17, rub:roz:mas:ara:zyc:bru:wal:22}. In addition, we make use of the dSD protocol as an illuminative example to apply the process matrix formalism to multipartite systems of identical particles.

The paper is organised as follows. In Section~\ref{Sec:StateOfTheArt} we give a short overview of the process matrix formalism and the dSD protocol. Section~\ref{Sec:ProcessMatrixDescription} is devoted to the process matrix formalism description of dSD protocol, and to the discussion of the {\em operational} role and importance of the vacuum state for its description. In particular, in Subsection~\ref{sec:analysis} we present the argument for our main conclusion, namely that the interaction with the vacuum should be considered an operation.  In Section~\ref{Sec:IdenticalParticles} we provide the basic rules for the application of the process matrix formalism to identical multiparticle systems. Section~\ref{Sec:Conclusions} is devoted to the summary of our results, discussion and prospects for future research. The Appendix contains various technical details of the calculations.

\section{\label{Sec:StateOfTheArt}State of the art}

In this section, we present an overview of the relevant background results. First, we give a short introduction to the process matrix formalism, and then we present the dSD protocol. This overview is not intended to be complete or self-contained, but merely of informative type. The reader should consult the literature for more details.

\subsection{\label{Sec:ProcessMatrix}The process matrix formalism}

The process matrix formalism is based on an idea of a set of laboratories, interacting with the outside world by exchanging quantum systems. Each laboratory is assumed to be spatially local in the sense that one can consider its size negligible for the problem under discussion. Inside the laboratory, it is assumed that the ordinary laws of quantum theory hold. The laboratory interacts with the outside world by receiving an {\em input} quantum system and by sending an {\em output} quantum system. Inside the laboratory, the input and output quantum systems are being manipulated using the notion of an {\em instrument}, denoting the most general operation one can perform over quantum systems. Each interaction is also assumed to be localised in time, such that each operation of a given laboratory has a separate spacetime point assigned to it (see Subsection~\ref{sec:analysis} for a discussion of time delocalised laboratories and operations~\cite{ore:19}). Thus, we introduce the notion of a {\em gate}, which represents the action of an instrument at a given spacetime point (see Section~2 of~\cite{pau:voj:20}); for simplicity, both the gate and its corresponding spacetime point will be denoted by the same symbol, $G$. By $G_I$ and $G_O$ we denote the Hilbert spaces of the input and the output quantum systems, respectively. These Hilbert spaces are assumed to be finite-dimensional or trivial. The action of the instrument is described by an operator, $\cM^G_{x,a}:G_I\otimes G_I^* \to G_O \otimes G_O^*$, which may depend on some classical input information $a$ available to the gate $G$, and some readouts $x$ of eventual measurement results that may take place in $G$. Thus, the instrument maps a generic mixed input state $\rho_I$ into the output state $\rho_O=\cM^G_{x,a}(\rho_I)$.

Given such a setup, one defines a {\em process}, denoted $\cW$, as a functional over the instruments of all gates, as
\begin{equation*}
p(x,y,\dots \vert a,b,\dots) = \cW ( \cM^{G^{(1)}}_{x,a} \otimes \cM^{G^{(2)}}_{y,b} \otimes \dots )\,,
\end{equation*}
where $p(x,y,\dots \vert a,b,\dots)$ represents the probability of obtaining measurement results $x,y,\dots $, given the inputs $a,b,\dots $. In order for the right-hand side to be interpreted as a probability distribution, the process $\cW$ must satisfy three basic axioms,
\begin{equation}
\label{eq:ProcessMatrixAxioms}
\begin{array}{rcl}
  \cW \!\!\!\! & \geq & \!\!\!\! 0 \vphantom{\ds\frac{A}{A}} \,, \\
  \ds \tr \cW  \!\!\!\! & = &  \!\!\!\! \ds \prod\nolimits_i^{\vphantom{A}} \dim G^{(i)}_O\,, \\
  \cW \!\!\!\! & = & \!\!\!\! \cP_G (\cW) \vphantom{\ds\frac{A}{A}}\,, \\
\end{array}
\end{equation}
where $\cP_G$ is a certain projector onto a subspace of $\bigotimes_i \left(G^{(i)}_I \otimes G^{(i)}_O\right)$ which, together with the second requirement, ensures the normalisation of the probability distribution (see~\cite{ara:bra:cos:fei:gia:bru:15} for details).

In order to have a computationally manageable formalism, one often employs the Choi-Jamio\l kowski (CJ) map over the instrument operations, such that a given operation $\cM^G_{x,a}$ is being represented by a matrix,
\begin{equation} 
  \label{eq:DefCJmap}
\begin{array}{rcl}
  M^G_{x,a} \!\!\!\! & = & \!\!\!\! \ds \Big[ \left( \cI \otimes \cM^G_{x,a} \right) \left( \kket{\one} \bbra{\one} \right) \Big]^T \\
  & \in & \!\!\!\! (G_I\otimes G_O) \otimes (G_I\otimes G_O)^*, \vphantom{\ds\int} \\
\end{array}
\end{equation}
where
\begin{equation} 
\label{eq:DefTransportVector}
\kket{\one} \equiv \sum_i \ket{i} \otimes \ket{i} \in G_I \otimes G_I
\end{equation}
is the so-called {\em transport vector}, representing the non-normalised maximally entangled state, and $\cI$ is the identity operator. Then, one can describe the process $\cW$ using the {\em process matrix} $W$ to write
\begin{equation} 
  \label{eq:VerovatnocaPrekoProcesMatrice}
\begin{array}{l}
  p(x,y,\dots \vert a,b,\dots) =  \vphantom{\ds\int} \\
 \hphantom{mmmm} \ds \tr \left[ ( M^{G_1}_{x,a} \otimes M^{G_2}_{y,b} \otimes \dots ) W \right] \, .  \vphantom{\ds\int} \\
\end{array}
\end{equation}
Finally, if an instrument $\cM^G_{x,a}$ is linear, one can also use a corresponding ``vector'' notation (see Appendix A.1 in~\cite{ara:bra:cos:fei:gia:bru:15}),
\begin{equation} 
\label{eq:DefGateActionVector}
\kket{(\cM^{G}_{x,a})^*} \equiv \left[ \cI \otimes (\cM^{G}_{x,a})^* \right] \kket{\one} \in G_I\otimes G_O\,,
\end{equation}
so that
\begin{equation}
\label{eq:DefMatrixPrekoGateVectora}
M^G_{x,a} = \kket{(\cM^{G}_{x,a})^*} \bbra{(\cM^{G}_{x,a})^*}\,.
\end{equation}
In cases where all instruments are linear, and in addition the process matrix $W$ is a one-dimensional projector, one can introduce the the corresponding {\em process vector} $\kket{W}$, such that $W = \kket{W} \bbra{W}$, and rewrite~(\ref{eq:VerovatnocaPrekoProcesMatrice}) in the form:
\begin{equation} 
\label{eq:DefProbabilityDistribution}
\begin{array}{c}
p(x,y,\dots \vert a,b,\dots) = \vphantom{\ds\int}\hphantom{mmmmmmmmmi}\\
\hphantom{}\left\Vert \left( \bbra{\cM^{G^{(1)}*}_{x,a}} \otimes \bbra{\cM^{G^{(2)}*}_{y,b}} \otimes \dots \right) \kket{W} \strut \right\Vert^2 \!\! . \\
\end{array}
\end{equation}

\subsection{\label{Sec:DdSprotocol}The \texorpdfstring{\lowercase{d}}{}SD protocol}

In a recent paper~\cite{dak:san:18}, del Santo and Daki\'c have introduced a protocol which allows two agents to guess each other's input bits with certainty by exchanging a single particle only once. The protocol goes as follows. Initially, a single particle is prepared in a superposition state of being sent to Alice and being sent to Bob. Upon receiving the particle, both Alice and Bob perform unitary operations on it, encoding their bits of information, $a$ and $b$, respectively, about the outcomes of their coin tosses. They do this by changing the local phase of the particle by $(-1)^a$ and $(-1)^b$. The particle is subsequently forwarded to a beam splitter, and after that again to Alice and Bob, who now measure the presence or absence of the particle. 

This way, the state of the particle stays in {\em coherent} superposition of different paths in a Mach-Zehnder interferometer. The interference of its paths gives rise to {\em deterministic} outcome that depends on the relative phase $e^{i\phi} = (-1)^{a\oplus b}$ between the two branches: in case $\phi=0$, the particle will end up in Alice's laboratory, while otherwise it will end in Bob's. Thus, knowing their own inputs and the outcomes of their local measurements, both agents can determine each other's inputs, allowing for two-way communication using only one particle. This is clearly impossible in classical physics, demonstrating yet another example of the advantage of quantum over classical strategies.

The crucial aspect of the protocol lies in the fact that the {\em absence} of the particle represents a useful piece of information for an agent. This gives rise to the notion of the {\em vacuum state} as a carrier of information, playing the central role in the protocol. Thus, in order to describe the protocol using the process matrix formalism, one has to incorporate the notion of the vacuum in the formalism itself. We show this in detail in the next section.

It is interesting to note that the crucial role of the vacuum plays an important part not only in the dSD protocol, but also in a completely different setup that has been discussed a lot in recent literature, namely the {\em quantum switch}~\cite{chi:dar:per:val:13}. As analysed in detail in~\cite{pau:voj:20}, if one takes care to distinguish the two temporal positions of a given laboratory and introduces the notion of a vacuum explicitly, one can demonstrate that the optical implementations of the quantum switch in flat spacetime do not feature any superposition of causal orders induced by the spacetime metric. Instead, it was argued that superpositions of spacetime causal orders can be present only within the context of a theory of quantum gravity. As we shall see below, the notion of the interaction with the vacuum will prove essential to the process matrix description of the dSD protocol as well.

\section{\label{Sec:ProcessMatrixDescription}Process matrix description of the \texorpdfstring{\lowercase{d}}{}SD protocol }

\subsection{\label{SubSec:Preliminaries}The spacetime diagram}

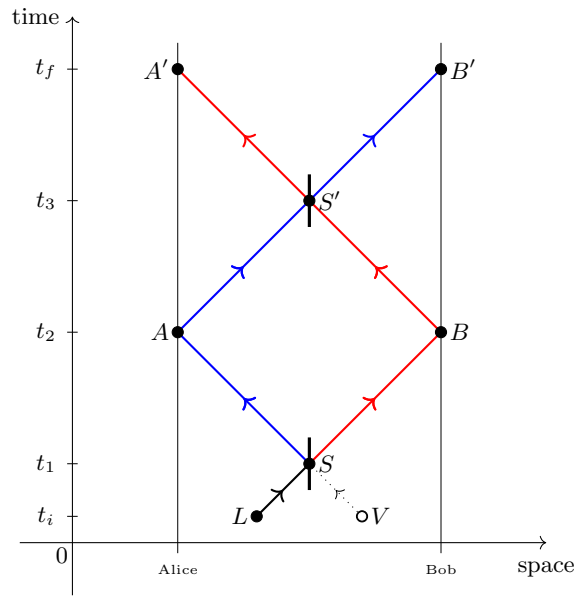
\begin{figure}[t]
\begin{center}
\begin{tikzpicture}[scale=0.7]

\tikzset{->-/.style={decoration={
  markings,
  mark=at position #1 with {\arrow{>}}},postaction={decorate}}}

\draw[->] (-1,-1) -- (9,-1);
\node at (9,-1.5) {space};

\draw[->] (0,-2) -- (0,9);
\node at (-0.7,9) {time};
\draw[very thin] (-0.1,-0.5) -- (0.1,-0.5);
\node at (-0.5,-0.5) {$t_i$};
\draw[very thin] (-0.1,0.5) -- (0.1,0.5);
\node at (-0.5,0.5) {$t_1$};
\draw[very thin] (-0.1,3) -- (0.1,3);
\node at (-0.5,3) {$t_2$};
\draw[very thin] (-0.1,5.5) -- (0.1,5.5);
\node at (-0.5,5.5) {$t_3$};
\draw[very thin] (-0.1,8) -- (0.1,8);
\node at (-0.5,8) {$t_f$};
\node at (-0.2,-1.25) {$0$};

\draw[very thin] (2,-1.2) -- (2,8.5);
\node at (2,-1.5) {\tiny Alice};

\draw[very thin] (7,-1.2) -- (7,8.5);
\node at (7,-1.5) {\tiny Bob};
 
\draw [->-=.5] [thick,black] (3.5,-0.5) -- (4.5,0.5);
\draw[->-=.5][dotted,black] (5.5,-0.5) -- (4.5,0.5);


\draw[->-=.25][->-=.75][thick,blue] (4.5,0.5) -- (2,3) -- (4.5,5.5);

\draw[->-=.25][->-=.75][thick,red] (4.5,0.5) -- (7,3) -- (4.5,5.5);


\draw[->-=.5][thick,red] (4.5,5.5) -- (2,8);

\draw[->-=.5][thick,blue] (4.5,5.5) -- (7,8);

\filldraw[black] (2,3) circle (3pt) node[anchor=east] {$A$};
\filldraw[black] (2,8) circle (3pt) node[anchor=east] {$A^\prime$};

\filldraw[black] (7,3) circle (3pt) node[anchor=west] {$B$};
\filldraw[black] (7,8) circle (3pt) node[anchor=west] {$B^\prime$};

\draw[very thick] (4.5,0) -- (4.5,1);
\draw[very thick] (4.5,5) -- (4.5,6);
\filldraw[black] (4.5,0.5) circle (3pt) node[anchor=west] {$S$};
\filldraw[black] (4.5,5.5) circle (3pt) node[anchor=west] {$S^\prime$};


\filldraw[black] (3.5,-0.5) circle (3pt) node[anchor=east] {$L$};
\filldraw[black] (5.5,-0.5) circle (3pt) node[anchor=west] {$V$};
\filldraw[white] (5.5,-0.5) circle (2pt);

\end{tikzpicture}
\caption{\label{SlikaJedan}The complete spacetime diagram of the process corresponding to the dSD protocol.}
\end{center}
\end{figure}

We begin by drawing the spacetime diagram of the process corresponding to the dSD protocol (see Figure~\ref{SlikaJedan}).

At the initial time $t_i$ the laser $L$ creates a photon and shoots it towards the beam splitter $S$, which at time $t_1$ performs the Hadamard operation and entangles it with the incoming vacuum state (described by the dotted arrow from the grey gate $V$). The entangled state of the photon and the vacuum continues on towards Alice's and Bob's gates $A$ and $B$, respectively. At time $t_2$, Alice and Bob generate their random bits $a$ and $b$, and encode them into the phase of the incoming photon-vacuum system. The system then proceeds to the beam splitter $S'$ which again performs the Hadamard operation at time $t_3$. The photon-vacuum system then proceeds to the gates $A'$ and $B'$, where it is measured at time $t_f$ by Alice and Bob, respectively. Note that the spatial distance $\Delta l$ between Alice and Bob is precisely equal to the time distance between the generation of the random bits and the final measurements,
\begin{equation*}
\Delta l = c( t_f - t_2 ) \, ,
\end{equation*}
so that a single photon has time to traverse the space between Alice and Bob {\em only once}. Also, note that the gate $V$, which generates the vacuum state, corresponds to a ``trivial instrument'', since the vacuum does not require any physical device to be generated. Nevertheless, the vacuum is still a legitimate physical state of the EM field, so the appropriate gate $V$ has to be formally introduced and accounted for in the process matrix formalism calculations.

\subsection{Formulation of the process matrix}

Based on the spacetime diagram, we formulate the process matrix description as follows. All spacetime points, where interaction between the EM field and some apparatus may happen, are assigned a gate and an operation which describes the interaction. Each gate has an input and output Hilbert space, as follows:
\begin{equation*}
\begin{array}{rlcl}
L: & \kompleksni & \to & L_O \, , \\
V: & \kompleksni & \to & V_O \, , \\
 & & & \\
S: & S_I & \to & S_O \, , \\
S': & S'_I & \to & S'_O \, , \\
\end{array}
\qquad
\begin{array}{rlcl}
A: & A_I & \to & A_O \, , \\
B: & B_I & \to & B_O \, , \\
 & & & \\
A': & A'_I & \to & \kompleksni \, , \\
B': & B'_I & \to & \kompleksni \, . \\
\end{array}
\end{equation*}
The initial gates, $L$ and $V$, have trivial input spaces and nontrivial output spaces. The final gates, $A'$ and $B'$, have trivial output spaces and nontrivial input spaces. The gates $A$ and $B$ have nontrivial input and output spaces. Each nontrivial space is isomorphic to $\cH_1\oplus \cH_0 \subset \cF$, where $\cH_0$ and $\cH_1$ are the vacuum and single-excitation subspaces of the Fock space $\cF$ in perturbative QED. Namely, by design of the dSD protocol, Alice and Bob may exchange at most one photon, which means that multiparticle Hilbert subspaces of the Fock space can be omitted. Moreover, the resulting probability distribution of the experiment outcomes does not in principle depend on the frequency or the polarisation of the photon in use, so we can approximate the single-excitation space as a one-dimensional Hilbert space. Given that the vacuum Hilbert space $\cH_0$ is by definition one-dimensional, we can write
\begin{equation*}
\cH_0 = \sspan \{ \ket{0} \} \equiv \kompleksni \,, \qquad
\cH_1 \approx \sspan \{ \ket{1} \} \equiv \kompleksni \,,
\end{equation*}
so that $\cH_0 \oplus \cH_1 \equiv \kompleksni \oplus \kompleksni$. Here, $\ket{0}$ and $\ket{1}$ denote the states of the vacuum and the photon in the occupation number basis of the Fock space. Therefore, we have
\begin{equation*}
\begin{array}{c}
  L_O \cong V_O \cong A_I \cong A_O \cong \hphantom{mmmmmmmmmm} \\
  B_I \cong B_O \cong A'_I \cong B'_I \cong \kompleksni \oplus \kompleksni \, . \\
\end{array}
\end{equation*}
Finally, the input and output spaces of beam splitters $S$ and $S'$ are ``doubled'', since a beam splitter operates over two inputs to produce two outputs. In particular,
\begin{equation*}
\begin{array}{lcl}
S_I & = & S_I^L \otimes S_I^V \, , \\
S_O & = & S_O^A \otimes S_O^B \, , \\
\end{array}
\qquad
\begin{array}{lcl}
S'_I & = & S_I^{\prime A} \otimes S_I^{\prime B} \, , \\
S'_O & = & S_O^{\prime A} \otimes S_O^{\prime B} \, , \\
\end{array}
\end{equation*}
where again
\begin{equation*}
\begin{array}{c}
  S_I^L \cong S_I^V \cong S_O^A \cong S_O^B \cong \hphantom{mmmmmmmmmm} \\
  S_I^{\prime A} \cong S_I^{\prime B} \cong S_O^{\prime A} \cong S_O^{\prime B} \cong \kompleksni \oplus \kompleksni \, . \\
\end{array}
\end{equation*}

With all relevant Hilbert spaces defined, we formulate the action of each gate, using the CJ map in the form~(\ref{eq:DefGateActionVector}). The gates $L$ and $V$ simply generate the photon and the vacuum,
\begin{equation} 
\label{eq:PreparationGates}
\kket{L^*}^{L_O} = \ket{1}^{L_O}\, , \qquad \kket{V^*}^{V_O} = \ket{0}^{V_O}\, ,
\end{equation}
where $*$ is the complex conjugation. The action of the beam splitters is
\begin{equation} 
\label{eq:GateS}
  \kket{S^*}^{S_IS_O} = \left[ I^{S_IS_I} \otimes (H^*)^{S_OS_I} \right] \kket{\one}^{S_IS_I}\, ,
\end{equation}
and
\begin{equation} 
\label{eq:GateSprime}
\kket{S^{\prime *}}^{S'_IS'_O} = \left[ I^{S'_IS'_I} \otimes (H^*)^{S'_OS'_I} \right] \kket{\one}^{S'_IS'_I}\, ,
\end{equation}
where the Hadamard operator for $S$ is defined as
\begin{equation*}
\begin{array}{l}
  H^{S_OS_I} = \\
 \hphantom{m} \ds \frac{1}{\sqrt{2}} \left( \ket{1}^{S_O^A} \ket{0}^{S_O^B} + \ket{0}^{S_O^A} \ket{1}^{S_O^B} \right) \bra{1}^{S_I^L} \bra{0}^{S_I^V} \\
 \hphantom{m} \ds + \frac{1}{\sqrt{2}} \left( \ket{1}^{S_O^A} \ket{0}^{S_O^B} - \ket{0}^{S_O^A} \ket{1}^{S_O^B} \right) \bra{0}^{S_I^L} \bra{1}^{S_I^V} , \\
\end{array}
\end{equation*}
and analogously for $H^{S'_OS'_I}$. The unit operator is denoted as $I$. Next, in the gates $A$ and $B$, Alice and Bob generate their random bits $a$ and $b$, and encode them into the phase of the photon. The corresponding actions are defined as
\begin{equation} 
\label{eq:GateA}
\kket{A^*}^{A_IA_O} = \left[ I^{A_IA_I} \otimes ( A^* )^{A_OA_I}  \right] \kket{\one}^{A_IA_I} \, ,
\end{equation}
and
\begin{equation} 
\label{eq:GateB}
\kket{B^*}^{B_IB_O} = \left[ I^{B_IB_I} \otimes ( B^* )^{B_OB_I}  \right] \kket{\one}^{B_IB_I}  ,
\end{equation}
where
\begin{equation*}
A^{A_OA_I} = (-1)^a \ket{1}^{A_O}\bra{1}^{A_I} \oplus \ket{0}^{A_O} \bra{0}^{A_I} \, ,
\end{equation*}
and
\begin{equation*}
B^{B_OB_I} = (-1)^b \ket{1}^{B_O}\bra{1}^{B_I} \oplus \ket{0}^{B_O} \bra{0}^{B_I} \, .
\end{equation*}
Finally, the gates $A'$ and $B'$ describe Alice's and Bob's measurement of the incoming state in the occupation number basis,
\begin{equation} 
\label{eq:MeasurementGates}
\kket{A^{\prime *}}^{A'_I} = \ket{a'}\, , \qquad \kket{B^{\prime *}}^{B'_I} = \ket{b'}\, ,
\end{equation}
where their respective measurement outcomes $a'$ and $b'$ take values from the set $\{0,1\}$, depending on whether the vacuum or the photon has been measured, respectively.

After specifying the actions of the gates, the last step is the construction of the process vector $\kket{W_{dSD}}$ itself. The dSD protocol assumes that the state of the photon remains unchanged during its travel between the gates. Therefore, the process vector will be a tensor product of transport vectors~(\ref{eq:DefTransportVector}), one for each line connecting two gates in the spacetime diagram. The input and output spaces of the gates connected by the line determine the spaces of the corresponding transport vector. Thus, the total process vector is:
\begin{equation} 
\label{eq:ProcessVectorDdS}
\begin{array}{l}
\kket{W_{dSD}} = \vphantom{\ds\int} \\
 \hphantom{m}\kket{\one}^{L_OS_I^L}
\kket{\one}^{V_OS_I^V} 
\kket{\one}^{S_O^AA_I}
\kket{\one}^{S_O^BB_I}  \vphantom{\ds\int} \\
 \hphantom{m} \kket{\one}^{A_OS_I^{\prime A}}
\kket{\one}^{B_OS_I^{\prime B}}
\kket{\one}^{S_O^{\prime A} A'_I}
\kket{\one}^{S_O^{\prime B} B'_I} \, . \vphantom{\ds\int} \\
\end{array}
\end{equation}

\subsection{Evaluation of the probability distribution}

Now that the process vector and the operations of all gates have been specified in detail, we can evaluate the probability distribution
\begin{equation} 
\label{eq:ProbabilityFromTheAmplitude}
p(a',b'|a,b) = \| \cM \|^2 \,,
\end{equation}
where the probability amplitude $\cM$ is obtained by taking the scalar product of $\kket{W_{dSD}}$ with the tensor product of all gates, see~(\ref{eq:DefProbabilityDistribution}). It is most instructive to perform the computation iteratively, taking the partial scalar product of $\kket{W_{dSD}}$ with each gate, one by one. The explicit calculation of each step is based on two lemmas from Appendix~\ref{AppA}.

We begin by taking the partial scalar product of~(\ref{eq:ProcessVectorDdS}) and the preparation gates~(\ref{eq:PreparationGates}). Using Lemma 1 from Appendix~\ref{AppA}, we obtain:
\begin{equation*}
\begin{array}{l}
\left( \bbra{L^*}^{L_O} \otimes \bbra{V^*}^{V_O} \right)\kket{W_{dSD}} = \\
\hphantom{mmmmm}\ket{1}^{S_I^L}
\ket{0}^{S_I^V} 
\kket{\one}^{S_O^AA_I}
\kket{\one}^{S_O^BB_I} \\
\hphantom{mmmmm}\kket{\one}^{A_OS_I^{\prime A}}
\kket{\one}^{B_OS_I^{\prime B}}
\kket{\one}^{S_O^{\prime A} A'_I}
\kket{\one}^{S_O^{\prime B} B'_I} \, . \\
\end{array}
\end{equation*}
Next we take the partial scalar product with the beam splitter S gate operation~(\ref{eq:GateS}). Using Lemma 2 from Appendix~\ref{AppA}, we obtain:
\begin{equation*}
\begin{array}{l}
\left( \bbra{S^*}^{S_IS_O} \otimes \bbra{L^*}^{L_O} \otimes \bbra{V^*}^{V_O} \right)\kket{W_{dSD}} = \\
\hphantom{mmmmm}
\ds \frac{1}{\sqrt{2}} \left( \ket{1}^{A_I} \ket{0}^{B_I} + \ket{0}^{A_I} \ket{1}^{B_I} \right) \\
\hphantom{mmmmm}
\kket{\one}^{A_OS_I^{\prime A}}
\kket{\one}^{B_OS_I^{\prime B}}
\kket{\one}^{S_O^{\prime A} A'_I}
\kket{\one}^{S_O^{\prime B} B'_I} \, . \\
\end{array}
\end{equation*}
Now we apply the Alice's gate operation~(\ref{eq:GateA}) to obtain:
\begin{equation*}
\begin{array}{l}
  \left( \bbra{A^*}^{A_IA_O} \otimes \bbra{S^*}^{S_IS_O} \otimes \right. \\
\hphantom{mm}\left. 
\bbra{L^*}^{L_O} \otimes \bbra{V^*}^{V_O} \right)\kket{W_{dSD}} = \\
\hphantom{mmmmm}
\ds \frac{1}{\sqrt{2}} \left( (-1)^a \ket{1}^{S_I^{\prime A}} \ket{0}^{B_I} + \ket{0}^{S_I^{\prime A}} \ket{1}^{B_I} \right) \\
\hphantom{mmmmm}
\kket{\one}^{B_OS_I^{\prime B}}
\kket{\one}^{S_O^{\prime A} A'_I}
\kket{\one}^{S_O^{\prime B} B'_I} \, . \\
\end{array}
\end{equation*}
Similarly, applying Bob's gate~(\ref{eq:GateB}) we get:
\begin{equation*}
\begin{array}{l}
  \left( \bbra{B^*}^{B_IB_O} \otimes \bbra{A^*}^{A_IA_O} \otimes \bbra{S^*}^{S_IS_O} \otimes \right. \\
\hphantom{mmmm}\left. \bbra{L^*}^{L_O} \otimes \bbra{V^*}^{V_O} \right)\kket{W_{dSD}} = \\
\hphantom{mm}
\ds \frac{1}{\sqrt{2}} \left( (-1)^a \ket{1}^{S_I^{\prime A}} \ket{0}^{S_I^{\prime B}} + (-1)^b \ket{0}^{S_I^{\prime A}} \ket{1}^{S_I^{\prime B}} \right) \\
\hphantom{mm}
\kket{\one}^{S_O^{\prime A} A'_I}
\kket{\one}^{S_O^{\prime B} B'_I} \, . \\
\end{array}
\end{equation*}
The next step is the application of the second beam splitter gate~(\ref{eq:GateSprime}). After a little bit of algebra, the result is:
\begin{widetext}
  \begin{equation*}
\begin{array}{l}
\left( \bbra{S^{\prime *}}^{S'_IS'_O} \otimes \bbra{B^*}^{B_IB_O} \otimes \bbra{A^*}^{A_IA_O} \otimes 
\bbra{S^*}^{S_IS_O} \otimes \bbra{L^*}^{L_O} \otimes \bbra{V^*}^{V_O} \right)\kket{W_{dSD}} = \vphantom{\ds\int} \\
\ds \hphantom{mmmmmmmmmmmmmmmmmmmmmm} \frac{(-1)^a + (-1)^b}{2} \ket{1}^{A'_I} \ket{0}^{B'_I} + \frac{(-1)^a - (-1)^b}{2} \ket{0}^{A'_I} \ket{1}^{B'_I} \, . \\
\end{array}
\end{equation*}
\end{widetext}
Finally, applying the measurement gates~(\ref{eq:MeasurementGates}), we obtain the complete probability amplitude,
\begin{equation*}
\begin{array}{lcl}
  \cM & = & \ds \frac{(-1)^a + (-1)^b}{2} \delta_{a'1} \delta_{b'0} \\
   & & \ds + \frac{(-1)^a - (-1)^b}{2} \delta_{a'0} \delta_{b'1} \, , \\
\end{array}
\end{equation*}
and substituting this into~(\ref{eq:ProbabilityFromTheAmplitude}), we obtain the desired probability distribution of the dSD process:
\begin{equation*}
\begin{array}{lcl}
p(a',b'|a,b) & = & \ds \frac{1 + (-1)^{a+b}}{2} \delta_{a'1} \delta_{b'0} \\    
 & & \ds + \frac{1 - (-1)^{a+b}}{2} \delta_{a'0} \delta_{b'1} \, . \\
\end{array}
\end{equation*}

From the probability distribution we can now conclude that there are two distinct possibilities: either Alice detects the photon and Bob does not, $a'=1,b'=0$, or vice versa, $a'=0,b'=1$. In the first case, because total probability must be equal to one, we have
\begin{equation*}
\frac{1 + (-1)^{a+b}}{2} =1\, , \qquad
\frac{1 - (-1)^{a+b}}{2} =0 \, . 
\end{equation*}
The only solution to these equations is $a=b$, which means that Alice and Bob have initially generated equal bits. Since both know the probability distribution and their own bit, they both know each other's bit as well, with certainty. In the second case, when Bob detects the photon, we instead have
\begin{equation*}
\frac{1 + (-1)^{a+b}}{2} =0\, , \qquad
\frac{1 - (-1)^{a+b}}{2} =1 \, ,
\end{equation*}
and the only solution is $a\neq b$, meaning that Alice and Bob have initially generated opposite bits. Again, both parties know the probability distribution and their own bit, and therefore each other's bit as well, with certainty.

In order to formalise this result, one can also introduce the parity $\pi \equiv a\oplus b$ and rewrite the probability distribution in the form
\begin{equation} 
\label{eq:DdSfullProbabilityDistribution}
\begin{array}{lcl}
  p(a',b'|\pi) & = & \ds \frac{1 + (-1)^{\pi}}{2} \delta_{a'1} \delta_{b'0} \\
  & & \ds + \frac{1 - (-1)^{\pi}}{2} \delta_{a'0} \delta_{b'1} \, . \\
\end{array}
\end{equation}
Thus, if Alice detects the photon, then $\pi$ is even, while if Bob detects the photon, $\pi$ must be odd. In both cases, they can ``guess'' each other's bits with certainty by~calculating
\begin{equation*}
x = \pi \oplus a \, , \qquad y = \pi \oplus b \, ,
\end{equation*}
where $x$ is Alice's prediction of the value of Bob's bit, and $y$ is Bob's prediction of the value of Alice's bit. Therefore, the probability of  guessing each other's input bit is
\begin{equation} 
\label{eq:ViolationOfGYNIresult}
p_{\text{success}} \equiv p(x=b \wedge y=a) = 1\, . 
\end{equation}

\subsection{Analysis of the process matrix description $\hphantom{mm}$ --- operational interpretation of the vacuum}\label{sec:analysis}

After we have given the detailed process matrix description of the dSD protocol and derived the result~(\ref{eq:ViolationOfGYNIresult}), we analyse in more detail the role of the vacuum in the formalism, giving its operational interpretation.

In order to clarify the exposition, let us give an overview of the argument, as follows:
\begin{itemize}
\item In the next paragraph below, we analyse the role of the vacuum in the dSD protocol, and conclude that the interaction with the vacuum should be regarded as an operation, on the same footing with all other interactions.
\item In the following four paragraphs, we discuss the optical implementation of the quantum switch protocol, which also features interactions between the agents and the vacuum. Since the same physical situation should always be described in the same way, we conclude that the interaction with the vacuum should be treated as an operation in this protocol as well. Thus, the protocol features a total of four, rather than two, operations.
\item Finally, in the remaining three paragraphs, we discuss the alternative point of view, namely that the interaction with the vacuum is not regarded as an operation. This is the case in the method for counting operations proposed in \cite{ara:cos:bru:14}. We conclude that it would then mean that in the dDS protocol an agent could extract information from the system at $t_f$ without performing an operation at all.

\end{itemize}

In the dSD protocol four operations (gates), $A, A^\prime, B$ and $B^\prime$ are performed (see Figure~\ref{SlikaJedan}). Note that for each choice of input bits $a$ and $b$ one of the two operations performed, $A^\prime$ and $B^\prime$, is of a special form: it represents the absence of the particle. This gives rise to an {\em operational interpretation}  of the vacuum state as a carrier of information, playing the central role in the protocol --- the very interaction between the apparatus and the vacuum (the absence of a particle) plays {\em exactly} the same role in this protocol as any other operation, i.e., not detecting a particle (``seeing the vacuum'') is an operation on its own. From the mathematical point of view, supported by the structure of the process vector~\eqref{eq:ProcessVectorDdS} that explicitly features the vacuum state, it is perfectly natural to consider the interaction between the apparatus  and the vacuum state on equal footing with the interaction between the apparatus and the field excitation (i.e., the particle). Both interactions equally represent operations. Therefore, one should regard the interaction with the vacuum as a resource, in the same way  as the interaction with the particle.

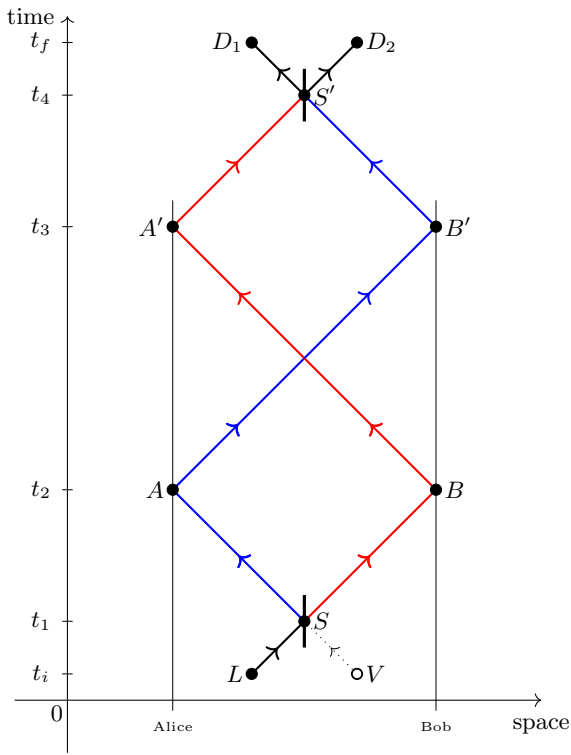
\begin{figure}[t]
\begin{center}
\begin{tikzpicture}[scale=0.7]

\tikzset{->-/.style={decoration={
  markings,
  mark=at position #1 with {\arrow{>}}},postaction={decorate}}}

\draw[->] (-1,-1) -- (9,-1);
\node at (9,-1.5) {space};

\draw[->] (0,-2) -- (0,12);
\node at (-0.7,12) {time};
\draw[very thin] (-0.1,-0.5) -- (0.1,-0.5);
\node at (-0.5,-0.5) {$t_i$};
\draw[very thin] (-0.1,0.5) -- (0.1,0.5);
\node at (-0.5,0.5) {$t_1$};
\draw[very thin] (-0.1,3) -- (0.1,3);
\node at (-0.5,3) {$t_2$};
\draw[very thin] (-0.1,8) -- (0.1,8);
\node at (-0.5,8) {$t_3$};
\draw[very thin] (-0.1,10.5) -- (0.1,10.5);
\node at (-0.5,10.5) {$t_4$};
\draw[very thin] (-0.1,11.5) -- (0.1,11.5);
\node at (-0.5,11.5) {$t_f$};
\node at (-0.2,-1.25) {$0$};

\draw[very thin] (2,-1.2) -- (2,8.5);
\node at (2,-1.5) {\tiny Alice};

\draw[very thin] (7,-1.2) -- (7,8.5);
\node at (7,-1.5) {\tiny Bob};
 
\draw [->-=.5] [thick,black] (3.5,-0.5) -- (4.5,0.5);
\draw[->-=.5][dotted,black] (5.5,-0.5) -- (4.5,0.5);


\draw[->-=.25][->-=.75][thick,blue] (4.5,0.5) -- (2,3) -- (4.5,5.5);

\draw[->-=.25][->-=.75][thick,red] (4.5,0.5) -- (7,3) -- (4.5,5.5);


\draw[->-=.5][thick,red] (4.5,5.5) -- (2,8);
\draw[->-=.5][thick,red] (2,8) -- (4.5,10.5);

\draw[->-=.5][thick,blue] (4.5,5.5) -- (7,8);
\draw[->-=.5][thick,blue] (7,8) -- (4.5,10.5);

\draw[->-=.5][thick,black] (4.5,10.5) -- (3.5,11.5);
\draw[->-=.5][thick,black] (4.5,10.5) -- (5.5,11.5);
\filldraw[black] (3.5,11.5) circle (3pt) node[anchor=east] {$D_1$};
\filldraw[black] (5.5,11.5) circle (3pt) node[anchor=west] {$D_2$};

\filldraw[black] (2,3) circle (3pt) node[anchor=east] {$A$};
\filldraw[black] (2,8) circle (3pt) node[anchor=east] {$A^\prime$};

\filldraw[black] (7,3) circle (3pt) node[anchor=west] {$B$};
\filldraw[black] (7,8) circle (3pt) node[anchor=west] {$B^\prime$};

\draw[very thick] (4.5,0) -- (4.5,1);
\draw[very thick] (4.5,10) -- (4.5,11);
\filldraw[black] (4.5,0.5) circle (3pt) node[anchor=west] {$S$};
\filldraw[black] (4.5,10.5) circle (3pt) node[anchor=west] {$S^\prime$};


\filldraw[black] (3.5,-0.5) circle (3pt) node[anchor=east] {$L$};
\filldraw[black] (5.5,-0.5) circle (3pt) node[anchor=west] {$V$};
\filldraw[white] (5.5,-0.5) circle (2pt);

\end{tikzpicture}
\caption{\label{SlikaDva}The complete spacetime diagram of the process corresponding to the optical quantum switch. Upon receiving the photon, Alice rotates its polarisation by the unitary $U$. Analogously, Bob performs rotation $V$ on the photon entering his lab.}
\end{center}
\end{figure}

Let us now consider the optical quantum switch, a similar protocol in which the notion of the vacuum also plays a role. Current optical implementations of the quantum switch feature four spacetime points, the same as the dSD protocol~\cite{pau:voj:20,vil:ren:22,orm:van:bar:22,vil:17}, thus having the similar type of the spacetime schematic description, see Figure~\ref{SlikaDva}. However, by introducing the notion of time delocalised operations it was argued that the optical switch implements only two operations, $U$ in spacetime points $A$ or $A^\prime$, and $V$ in spacetime points $B$ or $B^\prime$~\cite{ore:19}. Nevertheless, the optical switch features the same apparatus-vacuum interaction as the one from the dSD protocol: whenever the particle is in, say, the blue branch, and the operations $U$ and $V$ are applied at spacetime points $A$ and $B^\prime$, respectively, Alice's and Bob's labs experience the interaction with the vacuum at spacetime points $B$ and $A^\prime$ (and analogously for the red branch). Therefore, the treatment of the vacuum in the optical quantum switch is mutually incoherent with the treatment of the vacuum in the dSD protocol.

Our analysis can thus serve as motivation for a search towards a more coherent treatment of the vacuum  within the operational approach, since the same physical situation --- interaction between the apparatus and the vacuum --- is currently treated differently in the descriptions of the two protocols.

One might consider the following possible chain of inference. From the examples of both the quantum switch and the dSD protocol, we have that unitary operations (be it ``genuine rotations'' $U$ and $V$, as well as phase flips $\pm I$) are considered to be operations. From the example of the dSD protocol, we see that the interaction with the vacuum is an operation as well. Further, in reference~\cite{ore:19} it was argued that the optical switch features two ``time-delocalised operations'', $U$ and $V$. Thus, by the same token, it follows that within this operational approach the optical switch should feature two additional ``time-delocalised operations'': interactions with the vacuum, one performed by Alice, and the other by Bob (see Appendix \ref{AppB}). Therefore, the protocol features a total of four, rather than two, operations. Note that this is a possible treatment of the vacuum, which still features superposition of orders of operations $U$ and $V$ in the optical switch.

It is obvious that the interaction with the vacuum plays a prominent role in achieving the goal of the dSD protocol --- communication between Alice and Bob. But interactions with the vacuum are also crucial in the optical switch. Indeed, without those operations, it would be {\em impossible} to achieve superposition of orders of operations $U$ and $V$ in flat spacetime with fixed causal order of spacetime points~\cite{pau:voj:20}.

In~\cite{ara:cos:bru:14} the so-called ``flag'' systems were introduced to count the number of operations performed in a lab without destroying the superposition, which count only one operation per each lab of the optical switch. Note though that using this method, which effectively counts the number of times a particle enters the lab, one would count three rather than four operations in the dSD protocol. This means that either the method is not appropriate, or in fact the dSD protocol features three, instead of four operations. In the case of the former, it would be useful to introduce a formal operational definition of a general method of counting operations, given that the above ``flag'' method cannot count interactions with the vacuum. In the case of the latter, it would mean that one could extract the information from the system at $t_f$ without performing an operation at~all.

Indeed, if the interaction between the vacuum and the apparatus would not be considered an operation, an issue with formulating the process vector for the dSD protocol would arise. The one we formulated in~\eqref{eq:ProcessVectorDdS} contains input and output Hilbert spaces associated with the interaction between the vacuum and the detectors. It is not possible to formulate a process matrix for the dSD protocol that would feature three operations, without the mentioned interaction with the vacuum. Namely, depending on the choice of input bits $a$ and $b$, the photon will end up either in Alice's or Bob's lab, rendering it impossible to know in advance which of the two agents is supposed to perform the final operation. Thus, it is not possible to formulate a process matrix which features only one operation at the final moment $t_f$. Note that the process matrices themselves were introduced as the main tool for describing quantum processes in the operational approach. In other words, the impossibility of formulating the main operational tool for the dSD protocol without introducing the interaction with the vacuum as an operation, suggests that the latter should be considered as an operation in that protocol.

Note that, if the dSD and the optical switch protocols featured incoherent mixtures of the two possible paths instead of coherent superpositions, then one could formulate the corresponding process matrices without treating the interaction with the vacuum as an operation, indeed without even mentioning the vacuum at all. These would be purely classical processes, which would not feature any interference effects. In general, ommiting the vacuum is a natural point of view in classical physics. However, if one wants to describe quantum physics, the notions of the vacuum and its interaction with the apparatus are unavoidable.

\section{\label{Sec:IdenticalParticles}Identical particles}

The above analysis shows that the vacuum state plays a physically relevant role in transmitting information, and cannot be ignored. From the point of view of QFT this is a perfectly natural state of affairs, but from the point of view of quantum mechanics (QM) it is not, since the notion of vacuum as a physical state does not exist in QM a priori, and needs to be explicitly introduced by hand. Moreover, in QFT one can naturally study systems of indefinite number of identical particles. Therefore, as a first step towards the generalization of the process matrix formalism to QFT, we apply the existing abstract process matrix formalism to the representation of the second quantization.

In this section, we give basic elements of the process matrix formalism, when applied to systems of identical particles. In order to avoid working with (anti\=/)symmetrised vectors of multi-particle states that contain non-physical entanglement whenever two or more identical particles are fully distinguishable (say, one photon is in Alice's, and another in Bob's lab), we will use the representation of the second quantisation in which the effects of particle statistics are governed by the creation and annihilation (anti-)commutation rules. First, we need to move from the single-particle Hilbert spaces associated to the gates and the process matrix to the corresponding capped Fock spaces.

To each gate $G$, we  assign the input/output Fock spaces, $\cG_{I/O}$, given in terms of the vacuum state $\ket{0}$ and the single-particle Hilbert spaces $G_{I/O}$. The single-particle input Hilbert space is given as 
\begin{equation*}
G_I = \mbox{span} \{ \ket{i} = a_i^\dag \ket{0} \ | \ i= 1,2, \dots d_I\},
\end{equation*}
such that its creation and annihilation operators satisfy the standard (anti-)commutation relations,
\begin{equation} 
\label{eq:commutationrelations}
[a_i^\dag , a_j^\dag]_\pm = [a_i , a_j]_\pm = 0\,, \quad [a_i , a_j^\dag]_\pm = \delta_{ij}\,,	
\end{equation}
where $[\_\, ,\_]_+$ stands for anti-commutator, and $[\_\, ,\_]_-$ for commutator. The overall bosonic input Fock space is then
\begin{equation}
\label{eq:fock_input}
\cG_I = \bigoplus_{\ell=0}^\infty G_I(\ell),
\end{equation}
where $G_I(0) = \mbox{span} \{ \ket{0} \}$ is the zero-particle, $G_I(1) = G_I$ the single-particle, and
\begin{equation*}
G_I(\ell) = \{ [(a_1^\dag)^{s_1} ... (a_{d_I}^\dag)^{s_{d_I}}] \ket{0} \ | \ s_1 + ... + s_{d_I} = \ell\}
\end{equation*}
are the $\ell$-particle orthogonal subspaces of the input Fock space. For fermions, each $s_i \in \{ 0,1 \}$, and the orthogonal sum in Equation~\eqref{eq:fock_input} goes until $d_I$, instead of $\infty$. For a given gate, the output Fock space $\cG_O$ is defined analogously, and we denote its creation and annihilation operators as $\tilde{a}_i^\dag$ and $\tilde{a}_i$, respectively, in order to distinguish them from the corresponding operators in $\cG_I$.

Our formalism is constructed for quantum circuits which consist of finite number of gates. This means that we work in the approximation of a finite number of spacetime points, as opposed to the standard QFT where one works with an uncountably infinitely many spacetime points. Thus, given the algebra~(\ref{eq:commutationrelations}) for the creation and annihilation operators at a single gate, the full algebra across all gates is normalised to a Kronecker delta, instead of the standard Dirac delta function. Moreover, the operators in~(\ref{eq:commutationrelations}) are operators in coordinate space, as opposed to the momentum space operators which are standard in QFT, since they create and annihilate modes at a given gate (i.e., a given spacetime point), instead of modes with a given momentum. Taking into account our assumption of finite number of gates, the single-particle Hilbert spaces $G_{I/O}$ are finite-dimensional, i.e., $d_{I/O} \in \prirodni$. Since the gates are distinguishable, the modes assigned to different gates {\em always} (anti\=/)commute.

We restrict ourselves to the Minkowski spacetime, so that the global Poincar\'e symmetry implies that the vacuum state $\ket{0}$ is identical across different gates, as well as between input and output Fock spaces for a given gate. In this sense, each gate is assumed to be stationary in some inertial reference frame, since the Fock spaces of non-inertial gates would be subject to the Unruh effect. We leave the discussion of non-inertial gates and spacetimes with more general geometries for future work.

Once the Fock spaces have been defined, we pass on to the process matrix description of gate operations. Since a process matrix has to satisfy the normalisation rule (\ref{eq:ProcessMatrixAxioms}), the corresponding input and output spaces have to be finite-dimensional. To that end, we restrict ourselves to capped Fock spaces, which contain only a finite number of elements in the sum (\ref{eq:fock_input}), denoted $N\in\prirodni$. Together with the fact that $d_{I/O}$ is finite, it follows that the capped Fock spaces are finite-dimensional. A gate operation is represented via a CJ isomorphism of the corresponding operator between the input and the output capped Fock spaces, defined in equation~(\ref{eq:DefCJmap}),
\begin{equation}
\label{eq:DefCJmapDrugiPut}
M = \Big[ \left( \cI \otimes \cM \right) \left( \kket{\one} \bbra{\one} \right) \Big]^T \,,
\end{equation}
where the transport vector
\begin{equation} 
\label{eq:DefTransportVectorSpecial}
\kket{\one} = \sum_{k=0}^{N} \kket{\one_k}\,,
\end{equation}
is given in terms of $k$-transport vectors defined as
\begin{equation}
\label{eq:DefKdimTransportVectorSpecial}
\kket{\one_k}  = \sum \Big[\prod_{i=1}^d \frac{ (a_i^\dag)^{s_i}}{\sqrt{s_i!}}\Big] \otimes \Big[\prod_{i=1}^d \frac{(a_i^\dag)^{s_i}}{\sqrt{s_i!}}\Big] \ket{0}\,,
\end{equation}
where the sum is taken over all $s_i$ satisfying the constraint $s_1 + ... + s_d = k$.

One special case of the general formula (\ref{eq:DefCJmapDrugiPut}) is the case where gates destroy all coherence between $k$-particle sectors, for example by measuring the number of particles,
\begin{equation}
\label{eq:DefCJmapSpecialCaseJedan}
M = \sum_{k=0}^N \Big[ \left( \cI \otimes \cM_k \right) \left( \kket{\one_k} \bbra{\one_k} \right) \Big]^T \,,
\end{equation}
where $\cM_k$ represents the $k$-particle operator for the gate. The above gate represents a classical mixture of operations on each $k$-particle sector, as opposed to coherent superpositions of them.

Another special case of (\ref{eq:DefCJmapDrugiPut}), which does preserve the coherence between $k$-particle sectors, is represented by linear operations. For a linear gate operation, one can analogously use the ``vector'' formalism, and the generalisation of the CJ vector~(\ref{eq:DefGateActionVector}). With a slight abuse of notation, using $\cal M$ to denote the operator instead of its superoperator, we can now write
\begin{equation*}
\begin{array}{lcl}
  \kket{\cM^*} & = & \ds  \Big[\cI \otimes \cM^* \Big] \kket{\one} \\
  & = & \ds \sum_{k,k',k''=0}^N \Big[\cI_k \otimes \cM^*_{k'} \Big] \kket{\one_{k''}} \\
  & = & \ds \sum_{k=0}^N \Big[\cI_k \otimes \cM^*_k \Big] \kket{\one_k}\,, \\
\end{array}
\end{equation*}
since it is assumed that by definition
\begin{equation*}
\Big[ \cI_k \otimes \cM^*_{k'} \Big] \kket{\one_{k''}} \equiv 0\,, \qquad k'' \notin \{ k,k' \}\,.
\end{equation*}
Now, using (\ref{eq:DefMatrixPrekoGateVectora}) one can rewrite (\ref{eq:DefCJmapDrugiPut}) into the form
\begin{equation*}
\begin{array}{lcl}
M \!\!\! & = & \!\!\! \ds \kket{\cM^*} \bbra{\cM^*} \\
 & = &\!\!\! \ds \sum_{k,k'=0}^N \Big[\cI_k \otimes \cM^*_k \Big] \kket{\one_k} \bbra{\one_{k'}} \Big[\cI_{k'} \otimes \cM^*_{k'} \Big]^\dag\,,  \\
\end{array}
\end{equation*}
which is clearly different from the case (\ref{eq:DefCJmapSpecialCaseJedan}), since it contains off-diagonal elements which preserve coherence between $k$-particle sectors. One concrete example of this special case is the dSD protocol, discussed in the previous Section. Another example is a single-particle unitary operator
\begin{equation*}
U = \sum_{i,j} u_{ij} \tilde{a}_i^\dag a_j\,.
\end{equation*}
Then, its capped Fock-space generalisation is given as
\begin{equation*}
\cM = \sum_{k=0}^{N} \cM_k = \ket{0}\bra{0} + \sum_{k=1}^{N} \frac{1}{k!} : U^{\otimes k} : \,,
\end{equation*}
where $:U^{\otimes k}:$ is the normal ordering of $U^{\otimes k}$.

Given the capped Fock spaces and actions of instruments in all gates, a process matrix is defined in the same way as in Section~\ref{Sec:StateOfTheArt}, according to Eq.~(\ref{eq:VerovatnocaPrekoProcesMatrice}). A process matrix maps the tensor product of output spaces for all gates into the tensor product of input spaces for all gates. For example, if the process under consideration is a quantum circuit (see Section~2 of~\cite{pau:voj:20}), the corresponding process matrix can be represented as a tensor product of transport vectors, each corresponding to a wire connecting two gates. Transport vectors are defined in the same way as~(\ref{eq:DefTransportVectorSpecial}), where in~(\ref{eq:DefKdimTransportVectorSpecial}) the first set of creation operators corresponds to the input space of the wire, while the second set corresponds to its output space. Given that a wire is connecting two gates, its input and output spaces correspond to the output and input subspaces of the two gates, respectively. A gate can in general have multiple incoming or outgoing wires attached to it. Therefore, its input (output) space is a tensor product of all output (input) spaces of the corresponding wires.

\section{\label{Sec:Conclusions}Conclusions}

\subsection{Summary of the results}

In this work we have presented a detailed account of the dSD protocol, formulating it within the process matrix formalism. Analysing the role of the vacuum state in the dSD protocol and its process matrix description, we gave the operational interpretation of the vacuum. Our analysis shows that the interaction with the vacuum should be treated as an operation, on equal footing with all other interactions, thus representing a resource in quantum information protocols (including, for example, \cite{hsu:lai:cha:wu:lee:20, zha:che:chi:20}). As a consequence, the optical implementation of the quantum switch protocol features four rather than just two operations, in contrast to what was claimed in the literature \cite{ara:bra:cos:fei:gia:bru:15, bav:ara:bru:que:19, gue:rub:bru:18, pro:etal:15, rub:roz:fei:ara:zeu:pro:bru:wal:17, rub:roz:mas:ara:zyc:bru:wal:22}. Furthermore, we have applied the process matrix formalism to the second quantisation framework restricted to capped Fock spaces, providing the description of systems of identical particles.

\subsection{Discussion}

The first important point of this work is the necessity of explicitly introducing the interaction with the vacuum as a legitimate operation in the dSD protocol, on equal footing with any other operation. Indeed, the very lack of detection of the particle in the protocol provides an equal amount of information as its detection ({\em explicit} interaction). As a consequence, instead of interpreting the absence of particle as noninteraction, one should interpret it as the interaction between the vacuum and the apparatus, and thus as an operation. Including the interaction with the vacuum as an operation poses a question of the method of counting operations in a given protocol, since the operations corresponding to the interaction with the vacuum cannot be counted.

The introduction of the vacuum into the process matrix formalism gives a natural motivation to extend the latter to the case of identical particles, both bosons and fermions, which is the second important point of this work. However, note that while employing the formalism of second quantisation, our construction still features only a discrete number of gates. This discreteness means that we still work in {\em particle ontology} (i.e., mechanics). Nevertheless, our construction is an important first step towards defining the process matrix formalism in {\em field ontology}, i.e., fully fledged QFT.

\ \ %

\subsection{Future lines of investigation}

As mentioned in the discussion, a natural next line of investigation would be a generalisation of the process matrix formalism to full, or at least perturbative, QFT. This would include an analysis of non-inertial gates and the corresponding Unruh effect. In addition, a mathematically rigorous formulation of the axioms for the process matrix description in Fock spaces is also an important topic to be addressed. While the primary interest in process matrices lies in their application to higher order processes \cite{per:17,bis:per:19}, their generalisation to QFT would also be of great interest. Finally, addressing in more detail the interaction between the agent and the vacuum within the operational approach is an interesting topic of future research.

\bigskip

\centerline{\bf Acknowledgments}

\bigskip

The authors wish to thank Borivoje Daki\'c and Flavio del Santo for useful discussions.

RF acknowledges support from DP-PMI and Funda\c{c}\~ao para a Ci\^{e}ncia e Tecnologia (FCT) through Grant PD/BD/128636/2017, from FCT/MCTES through national funds and when applicable EU funds under the project UIDB/50008/2020, and the QuantaGENOMICS project, through the EU H2020 QuantERA II Programme, Grant Agreement No 101017733.

NP's work was partially supported by SQIG -- Security and Quantum Information Group of Instituto de Telecomunica\c{c}\~oes, by Programme (COMPETE 2020) of the Portugal 2020 framework [Project Q.DOT with Nr.\ 039728 (POCI-01-0247-FEDER-039728)] and the Funda\c{c}\~ao para a Ci\^{e}ncia e a Tecnologia (FCT) through national funds, by FEDER, COMPETE 2020, and by Regional Operational Program of Lisbon, under UIDB/50008/2020 (actions QuRUNNER, QUESTS), Projects QuantumMining POCI-01-0145-FEDER-031826, PRE\-DICT PTDC/CCI-CIF/29877/2017, CERN/FIS-PAR/0023/2019, QuantumPrime PTDC/EEI-TEL/8017/2020, as well as the FCT Est\'{i}mulo ao Emprego Cient\'{i}fico grant no. CEECIND/04594/2017/CP1393/CT000.

MV was supported by the Ministry of Science, Technological Development and Innovations of the Republic of Serbia, by the bilateral scientific cooperation between Portugal and Serbia through the project ``Symmetries and Quantization - 2020-2022'', no. 337-00-00227/2019-09/57 supported by the Portuguese Foundation for Science and Technology (FCT), Portugal, and the Ministry of Education, Science and Technological Development of the Republic of Serbia, and by the Science Fund of the Republic of Serbia, grant 7745968, ``Quantum Gravity from Higher Gauge Theory 2021'' --- QGHG-2021. The contents of this publication are the sole responsibility of the authors and can in no way be taken to reflect the views of the Science Fund of the Republic of Serbia.

\bibliographystyle{quantum}
\bibliography{vac-proc-mat}

\onecolumn
\appendix

\section{\label{AppA}Two lemmas for the process matrix evaluation}

\textbf{Lemma 1.} Let $\kket{\Psi^*}^{X_O} = \ket{\Psi^*}^{X_O}$ represent a gate which has no input, while it prepares the state $\ket{\Psi} \in X_O$ as its output. Then, the scalar product of that vector and the transport vector $\kket{\one}^{X_OY_I}$ is given as:
\begin{equation*}
{}^{X_O}\bbrakket{\Psi^*}{\one}^{X_OY_I} = \ket{\Psi}^{Y_I}\, .
\end{equation*}
\textit{Proof.} Using the fact that the transport vector is an unnormalized maximally entangled state, the explicit calculation goes as follows:
\begin{equation*}
\begin{array}{lcl}
  {}^{X_O}\bbrakket{\Psi^*}{\one}^{X_OY_I} & = & \ds \bra{\Psi^*}^{X_O} \sum_k \ket{k}^{X_O} \ket{k}^{Y_I} \\
  & = & \ds \sum_k \bracket{\Psi^*}{k} \; \ket{k}^{Y_I} \\
  & = & \ds \sum_k \bracket{k}{\Psi} \; \ket{k}^{Y_I} \\
  & = & \ds \ket{\Psi}^{Y_I} \, , \\
\end{array}
\end{equation*}
where we have used the unit decomposition $\ds I = \sum_k \ket{k} \bra{k}$ and the fact that $\bracket{\Psi^*}{k} = \bracket{\Psi}{k}^* = \bracket{k}{\Psi}$.

\medskip

\textbf{Lemma 2.} Let
\begin{equation*}
\kket{U^*}^{X_IX_O} = \left[ I^{X_IX_I} \otimes (U^*)^{X_OX_I} \right] \kket{\one}^{X_IX_I}
\end{equation*}
represent a gate which performs the operation $U : X_I \to X_O$, and let $\kket{W} = \ket{\Psi}^{X_I} \kket{\one}^{X_OY_I}$. Then the scalar product of the two is
\begin{equation*}
{}^{X_IX_O}\bbrakket{U^*}{W} = \Big( U\ket{\Psi} \Big)^{Y_I} \, .
\end{equation*}
\textit{Proof.} Again using the expansion of the transport vectors as unnormalized maximally entangled states, the explicit calculation goes as follows:
\begin{equation*}
\begin{array}{lcl}
{}^{X_IX_O}\bbrakket{U^*}{W} & = & \ds  \bbra{\one}^{X_IX_I}  \left[ I^{X_IX_I} \otimes (U^T)^{X_IX_O} \right] \ket{\Psi}^{X_I} \kket{\one}^{X_OY_I} \vphantom{\ds\sum_k} \\
 & = & \ds \sum_k \bra{k}^{X_I} \bra{k}^{X_I}  \left[ I^{X_IX_I} \otimes (U^T)^{X_IX_O} \right] \ket{\Psi}^{X_I} \sum_m \ket{m}^{X_O} \ket{m}^{Y_I} \\
 & = & \ds \sum_{k,m} \Big( \bra{k}^{X_I} I^{X_IX_I} \ket{\Psi}^{X_I} \Big) \Big( \bra{k}^{X_I} (U^T)^{X_IX_O} \ket{m}^{X_O} \Big) \ket{m}^{Y_I} \\
 & = & \ds \sum_{k,m} \Big( \bracket{k}{\Psi} \Big) \Big( \bra{m} U \ket{k} \Big) \ket{m}^{Y_I} \\
 & = & \ds \sum_m \bra{m} U \left( \sum_k \ket{k} \bra{k} \right) \ket{\Psi} \;  \ket{m}^{Y_I} \\
 & = & \ds \sum_m \bra{m} U \ket{\Psi} \;  \ket{m}^{Y_I} \\
 & = & \ds \Big( U \ket{\Psi} \Big)^{Y_I} \, , \\
\end{array}
\end{equation*}
where we have again used the unit decomposition and the fact that $\bra{k} U^T \ket{m} = \bra{m} U \ket{k}$.

\section{\label{AppB}Time-delocalised operations in the optical switch}

\begin{figure}[t]
\begin{center}
\begin{tabular}{ccc}
\begin{tikzpicture}[scale=0.5]

\tikzset{->-/.style={decoration={
  markings,
  mark=at position #1 with {\arrow{>}}},postaction={decorate}}}

\draw[->] (-1,-1) -- (9,-1);
\node at (9,-1.5) {space};

\draw[->] (0,-2) -- (0,12);
\node at (-0.7,12.2) {time};
\draw[very thin] (-0.1,-0.5) -- (0.1,-0.5);
\node at (-0.5,-0.5) {$t_i$};
\draw[very thin] (-0.1,0.5) -- (0.1,0.5);
\node at (-0.5,0.5) {$t_1$};
\draw[very thin] (-0.1,3) -- (0.1,3);
\node at (-0.5,3) {$t_2$};
\draw[very thin] (-0.1,8) -- (0.1,8);
\node at (-0.5,8) {$t_3$};
\draw[very thin] (-0.1,10.5) -- (0.1,10.5);
\node at (-0.5,10.5) {$t_4$};
\draw[very thin] (-0.1,11.5) -- (0.1,11.5);
\node at (-0.5,11.5) {$t_f$};
\node at (-0.2,-1.25) {$0$};

\draw[very thin] (2,-1.2) -- (2,8.5);
\node at (2,-1.5) {\tiny Alice};

\draw[very thin] (7,-1.2) -- (7,8.5);
\node at (7,-1.5) {\tiny Bob};
 
\draw [->-=.5] [thick,black] (3.5,-0.5) -- (4.5,0.5);
\draw[->-=.5][dotted,black] (5.5,-0.5) -- (4.5,0.5);


\draw[->-=.25][->-=.75][thick,blue] (4.5,0.5) -- (2,3) -- (4.5,5.5);

\draw[->-=.25][->-=.75][dotted,black] (4.5,0.5) -- (7,3) -- (4.5,5.5);


\draw[->-=.5][dotted,black] (4.5,5.5) -- (2,8);
\draw[->-=.5][dotted,black] (2,8) -- (4.5,10.5);

\draw[->-=.5][thick,blue] (4.5,5.5) -- (7,8);
\draw[->-=.5][thick,blue] (7,8) -- (4.5,10.5);

\draw[->-=.5][thick,black] (4.5,10.5) -- (3.5,11.5);
\draw[->-=.5][thick,black] (4.5,10.5) -- (5.5,11.5);
\filldraw[black] (3.5,11.5) circle (3pt) node[anchor=east] {$D_1$};
\filldraw[black] (5.5,11.5) circle (3pt) node[anchor=west] {$D_2$};

\filldraw[black] (2,3) circle (3pt) node[anchor=east] {$A$};
\filldraw[black] (2,8) circle (3pt) node[anchor=east] {$A^\prime$};
\filldraw[white] (2,8) circle (2pt);

\filldraw[black] (7,3) circle (3pt) node[anchor=west] {$B$};
\filldraw[white] (7,3) circle (2pt);
\filldraw[black] (7,8) circle (3pt) node[anchor=west] {$B^\prime$};

\draw[very thick] (4.5,0) -- (4.5,1);
\draw[very thick] (4.5,10) -- (4.5,11);
\filldraw[black] (4.5,0.5) circle (3pt) node[anchor=west] {$S$};
\filldraw[black] (4.5,10.5) circle (3pt) node[anchor=west] {$S^\prime$};


\filldraw[black] (3.5,-0.5) circle (3pt) node[anchor=east] {$L$};
\filldraw[black] (5.5,-0.5) circle (3pt) node[anchor=west] {$V$};
\filldraw[white] (5.5,-0.5) circle (2pt);

\end{tikzpicture}
& &
\begin{tikzpicture}[scale=0.5]

\tikzset{->-/.style={decoration={
  markings,
  mark=at position #1 with {\arrow{>}}},postaction={decorate}}}

\draw[->] (-1,-1) -- (9,-1);
\node at (9,-1.5) {space};

\draw[->] (0,-2) -- (0,12);
\node at (-0.7,12.2) {time};
\draw[very thin] (-0.1,-0.5) -- (0.1,-0.5);
\node at (-0.5,-0.5) {$t_i$};
\draw[very thin] (-0.1,0.5) -- (0.1,0.5);
\node at (-0.5,0.5) {$t_1$};
\draw[very thin] (-0.1,3) -- (0.1,3);
\node at (-0.5,3) {$t_2$};
\draw[very thin] (-0.1,8) -- (0.1,8);
\node at (-0.5,8) {$t_3$};
\draw[very thin] (-0.1,10.5) -- (0.1,10.5);
\node at (-0.5,10.5) {$t_4$};
\draw[very thin] (-0.1,11.5) -- (0.1,11.5);
\node at (-0.5,11.5) {$t_f$};
\node at (-0.2,-1.25) {$0$};

\draw[very thin] (2,-1.2) -- (2,8.5);
\node at (2,-1.5) {\tiny Alice};

\draw[very thin] (7,-1.2) -- (7,8.5);
\node at (7,-1.5) {\tiny Bob};
 
\draw [->-=.5] [thick,black] (3.5,-0.5) -- (4.5,0.5);
\draw[->-=.5][dotted,black] (5.5,-0.5) -- (4.5,0.5);


\draw[->-=.25][->-=.75][dotted,black] (4.5,0.5) -- (2,3) -- (4.5,5.5);

\draw[->-=.25][->-=.75][thick,red] (4.5,0.5) -- (7,3) -- (4.5,5.5);


\draw[->-=.5][thick,red] (4.5,5.5) -- (2,8);
\draw[->-=.5][thick,red] (2,8) -- (4.5,10.5);

\draw[->-=.5][dotted,black] (4.5,5.5) -- (7,8);
\draw[->-=.5][dotted,black] (7,8) -- (4.5,10.5);

\draw[->-=.5][thick,black] (4.5,10.5) -- (3.5,11.5);
\draw[->-=.5][thick,black] (4.5,10.5) -- (5.5,11.5);
\filldraw[black] (3.5,11.5) circle (3pt) node[anchor=east] {$D_1$};
\filldraw[black] (5.5,11.5) circle (3pt) node[anchor=west] {$D_2$};

\filldraw[black] (2,3) circle (3pt) node[anchor=east] {$A$};
\filldraw[white] (2,3) circle (2pt);
\filldraw[black] (2,8) circle (3pt) node[anchor=east] {$A^\prime$};

\filldraw[black] (7,3) circle (3pt) node[anchor=west] {$B$};
\filldraw[black] (7,8) circle (3pt) node[anchor=west] {$B^\prime$};
\filldraw[white] (7,8) circle (2pt);

\draw[very thick] (4.5,0) -- (4.5,1);
\draw[very thick] (4.5,10) -- (4.5,11);
\filldraw[black] (4.5,0.5) circle (3pt) node[anchor=west] {$S$};
\filldraw[black] (4.5,10.5) circle (3pt) node[anchor=west] {$S^\prime$};


\filldraw[black] (3.5,-0.5) circle (3pt) node[anchor=east] {$L$};
\filldraw[black] (5.5,-0.5) circle (3pt) node[anchor=west] {$V$};
\filldraw[white] (5.5,-0.5) circle (2pt);

\end{tikzpicture}
\end{tabular}
\caption{\label{SlikaTri}Two branches of the coherent superposition of the optical switch protocol.}
\end{center}
\end{figure}
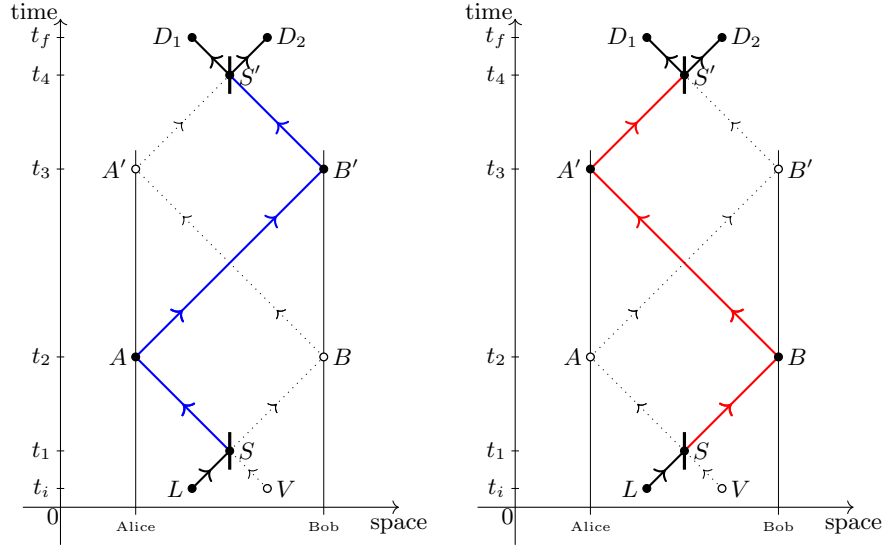

Figure~\ref{SlikaTri} depicts two branches coherently superposed in the optical switch. The left diagram represents the branch in which the photon first enters Alice's lab, and then Bob's. On the right, the photon first visits Bob's lab, and then Alice's. Whenever the photon enters Alice's lab, she applies unitary $U$ (in $A$, left diagram, or $A^\prime$, right diagram), while Bob interacts with the vacuum (in $B$, left diagram, or $B^\prime$, right diagram). Analogously, whenever the photon enters Bob's lab, he applies unitary $V$ (in $B$, right diagram, or $B^\prime$, left diagram), while Alice interacts with the vacuum (in $A$, right diagram, or $A^\prime$, left diagram).

Since applying the unitaries in a quantum protocol are operations, and since in the optical switch they are applied by Alice ($U$) and Bob ($V$) at two different times, we say that the optical switch features two time-delocalised operations $U$ (at $A$ and $A^\prime$) and $V$ (at $B$ and $B^\prime$).

Since the interaction with the vacuum in the dSD protocol is an operation, and since in the optical switch it is applied by Alice and Bob at two different times, one can say that the optical switch features two time-delocalised operations of the interaction with the vacuum (at $A$ and $A^\prime$, as well as at $B$ and~$B^\prime$).

\end{document}